# Mining Software Metrics from Jazz

Jacqui Finlay

Auckland University of Technology
Private Bag 92006, Wellesley Street
Auckland 1142
+64 9 921 9999

jacqui.finlay@aut.ac.nz

Dr Andy M. Connor

Auckland University of Technology
Private Bag 92006, Wellesley Street
Auckland 1142
+64 9 921 9999

andrew.connor@aut.ac.nz

Dr Russel Pears

Auckland University of Technology
Private Bag 92006, Wellesley Street
Auckland 1142
+64 9 921 9999

russel.pears@aut.ac.nz

## ABSTRACT

In this paper, we describe the extraction of source code metrics from the Jazz repository and the application of data mining techniques to identify the most useful of those metrics for predicting the success or failure of an attempt to construct a working instance of the software product. We present results from a systematic study using the J48 classification method. The results indicate that only a relatively small number of the available software metrics that we considered have any significance for predicting the outcome of a build. These significant metrics are discussed and implication of the results discussed, particularly the relative difficulty of being able to predict failed build attempts.

## Categories and Subject Descriptors

D.2.8 [**Metrics**]: Complexity Measures and Product Metrics.

## 1. INTRODUCTION

Software development projects involve the use of a wide range of tools to produce a software artifact. Software repositories such as source control systems have become a focus for emergent research as being a source of rich information regarding software development projects. The mining of such repositories is becoming increasingly common with a view to gaining a deeper understanding of the development process and building better prediction and recommendation systems. The Jazz development environment has been recognized as offering new opportunities in this area because it integrates the software archive and bug database by linking bug reports and source code changes with each other [1]. Whilst this provides much potential in gaining valuable insights into the development process of software projects, such potential is yet to be fully realized.

In this paper we describe an initial attempt to fully extract the richness available in the Jazz dataset by utilizing source code metrics as a means of directly measuring the impact of code issues on build success. A build is defined as an attempt to construct a working instance of the software product for which the metrics have been extracted, in this case the Jazz product. In the next section we provide a brief overview of related work. Section 3 discusses the nature of the Jazz data repository and metrics that we utilized to mine the repository. In section 4, we discuss our approach to mining the software repository in Jazz, while our initial results are presented in section 5. Finally, we conclude our paper with a discussion of the limitations of the current work and a plan for addressing these issues in future work.

## 2. BACKGROUND & RELATED WORK

According to Herzig & Zeller [1], Jazz offers not only huge opportunities for software repository mining but also a number of challenges. One of the opportunities is that Jazz provides a more detailed dataset in which all artifacts are linked to each other. To date, much of the work that utilizes Jazz as a repository has focused on the convenience provided by linking artifacts such as bug reports to specification items along with the team communication history. Researchers have focused on areas such as whether there is an association between team communication and build failure [2] or whether it is possible to identify relationships among requirements, people and software defects [3]. Other work [4] has focused purely on the collaborative nature of software development. To date, most of the work involving the Jazz dataset has focused on aspects other than analysis of the source code contained in the repository.

Research that focuses on the analysis of metrics derived from source code analysis to predict software defects has generally shown that there is no single code or churn metric capable of predicting failures [5, 6, 7], though evidence suggests that a combination can be used effectively[8]. To date no such source code analysis has been conducted on the Jazz project data and it is our goal to perform an in-depth analysis of the repository to gain insight into the usefulness of software product metrics in predicting software build failure.

Buse and Zimmerman [9] suggests that whilst software projects can be rated by a range of metrics that describe the complexity, maintainability, readability, failure propensity and many other important aspects of software development process health, it still continues to be risky and unpredictable. In their paradigm of software analytics, Buse and Zimmerman suggest that metrics themselves need to be utilised to gain insights and as such it is necessary to distinguish questions of information which some tools already provide (e.g., how many bugs are in the bug database?) from questions of insight which provide managers with an understanding of a project's dynamics (e.g., will the project be delayed?). They continue by suggesting that the primary goal of software analytics is to help managers move beyond information and toward insight, though this requires knowledge of the domain coupled with the ability to identify patterns involving multiple indicators.

It is our belief that the Jazz dataset provides sufficiently rich information to support these goals. In our work to date we have analysed the software product metrics available through Jazz and are in the process of exploring how such product metrics can be combined with process metrics to support the goals of the software analytics paradigm.

## 3. THE JAZZ DATASET

### 3.1 Overview of Jazz

IBM Jazz is a fully integrated software development tool that automatically captures software development processes and artifacts. The Jazz repository contains real-time evidence that allows researchers to gain insights into team collaboration and development activities within software engineering projects [10]. With Jazz it is possible to extract the interactions between contributors in a development project and examine the artifacts produced. This means that Jazz provides the capability to extract social network data and relate such data to the software project outcomes. Figure 1 illustrates that through the use of Jazz it is possible to visualize members, work items and project team areas.

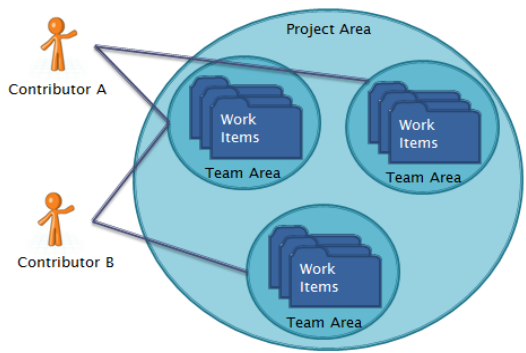

**Figure 1. Jazz Repository: Contributors, Project Area, Team Areas and Work Items.**

The Jazz repository artifacts include work items, build items, change sets, source code files, authors and comments. A work item is a description of a unit of work, which is categorized as a task, enhancement or defect. A build item is compiled software to form a working unit. A change set is a collection of code changes in a number of files. In Jazz a change set is created by one author only and relates to one work item. A single work item may contain many change sets. Source code files are included in change sets and over time can be related to multiple change sets. Authors are contributors to the Jazz project. Comments are recorded communication between contributors of a work item. Comments on work items are the primary method of information transfer among developers.

There are limitations for incorporating the Jazz repository into research. Firstly, the repository is highly complex and has huge storage requirements for tracking software artifacts. Another issue is that the repository contains holes and misleading elements which cannot be removed or identified easily. This is because the Jazz environment has been used within the development of itself; therefore many features provided by Jazz were not implemented at early stages of the project. We acknowledge the challenge in dealing with such inconsistency and are proposing an approach that delves further down the artifact chain than most previous work using Jazz. It is our premise that the early software releases were functional, so whilst the project "meta-data" may be missing details (such as developer comments) the source code should represent a stable system that can be analyzed to gain insight regarding the development project.

### 3.2 Software Metrics

Software metrics have been generated in order to deal with the sparseness of the data. Metric values can be derived from extracting development code from software repositories. Such metrics are commonly used within model-based project management methods. Software metrics are used to measure the complexity, quality and effort of a software development project [11]. The Jazz database contains over 200 relations, containing numerous cryptic fields. Thus data extraction via SQL queries runs the high risk of retrieving unreliable or incomplete data. Instead, we used the Jazz client/server APIs, an approach recommended in a study by Nguyen, Schröter, and Damian [10].

The Jazz repository consists of various types of software builds. Included in this study were continuous builds (regular user builds), nightly builds (incorporating changes from the local site) and integration builds (integrating components from remote sites). Source code files were extracted for each available build within the repository. Subsequently software metrics were generated by utilizing the IBM Rational Software Analyzer tool. As a result the following traditional, object orientated and Halstead software metrics were derived from the source code files for each build. These are listed in the following sections with each metric assigned a unique ID (the number in parentheses).

#### 3.2.1 Traditional Metrics

Traditional metrics included in our analysis are: *Number of attributes (1), Average number of attributes per class (2), Average number of constructors per class (3), Average number of comments (4), Average lines of code per method (5), Average number of methods (6), Average number of parameters (7), Number of types per package (8), Comment/Code Ratio (9), Number of constructors (10), Number of import statements (11), Number of interfaces (12), Lines of code (13).*

#### 3.2.2 Object Oriented Metrics

Object Orientated metrics included in our analysis are: *Number of comments (14), Number of methods (15), Number of parameters (16), Number of lines (17), Abstractness (18), Afferent coupling (19), Efferent coupling (20), Instability (21), Normalized Distance (22), Average block depth (23), Weighted methods per class (24), Maintainability index (25), Cyclomatic complexity (26), Lack of cohesion 1 (27), Lack of cohesion 2 (28), Lack of cohesion 3 (29).*

#### 3.2.3 Halstead Metrics

Halstead metrics included in our analysis are: *Number of operands (30), Number of operators (31), Number of unique operands (32), Number of unique operators (33), Number of delivered bugs (34), Difficulty level (35), Effort to implement (36), Time to implement (37), Program length (38), Program level (39), Program vocabulary size (40), Program volume (41), Depth of Inheritance (42).*

#### 3.2.4 Jazz Metrics

In addition to software (source code) metrics the following metrics were included to provide additional relevant data for each build. Whilst a range of metrics that are unique to the Jazz environment are available, at present this research only includes whether the build attempt is successful or whether it fails. A failed build is in essence one where the end product does not pass all of the test cases or does not behave as expected.

## 4. EXPERIMENTAL METHOD

This work revolves around the use of classification methods for the analysis of software metrics. For this purpose the Weka [12] machine learning workbench was used. There are various challenges that arise when adopting data mining approaches. Real life data is not always suitable for the mining process. There is often noise within the data, missing data, or even misleading data that can have negative impacts on the mining and learning process [13]. The project data that is extracted from Jazz was gathered during the development of Jazz. As a consequence features that automatically capture project processes did not exist until later development stages of Jazz (gaps would often appear at early stages of the project data set). Excluded from the data set were instances that had no work items associated with a build, build warning results and builds that had missing values within the derived software metrics.

Software metrics from continuous builds were used to construct the data set, however in doing so there were more instances of successful builds than failed builds. In order to balance the data set failed builds were injected from nightly and integration builds. This option was preferred over removing successful builds from the data set, thus decreasing the possibility of model over-fitting. In total, 129 builds were included, out of which there were 51 successful builds and 78 failed builds. This presents a situation where the number of features is fairly close to the number of instances available for analysis, which is not an ideal scenario from a Data Mining perspective. One possible solution was to increase the number of instances by including more builds but more data was not forthcoming from IBM at the time that the research was executed. We thus opted to alleviate this problem by using feature selection methods prior to classifying the data.

### 4.1 Dataset Interpretations

In the Jazz dataset a given build consists of a number of different work items. Each work item contains a *changeset* that indicates the actual source code files that are modified during the implementation of the work item. Source code metrics for each file calculated for each source code file using the IBM Software Analyser tool. In our work to date we have explored a number of different options for aggregating the metric values to produce a representative metric value for the complete build. Table 1 indicates the different mechanisms used for calculating these representative values.

**Table 1. Sample Datasets**

| Dataset ID | Description |
|---|---|
| 1 | The value for each metric for each source code file is calculated individually. The average value of the metric is propagated up to the build level |
| 2 | The value for each metric for each source code file is calculated individually. The maximum value of the metric is propagated up to the build level |
| 3 | The value for each metric for each source code file is calculated individually. The total sum of all metrics is propagated up to the build level |

For each dataset, a number of sub-datasets were considered where the number of features/metrics used was reduced. The reduced datasets are identified in the results are identified by the addition of a suffix which indicates one of the following:

a. Only traditional software metrics are used
b. Only object oriented metrics are used
c. Only Halstead metrics are used
d. The "average number of…" metrics are not used

In the results presented in the next section, a dataset ID of 2c would therefore indicate that maximum metric values for only the Halstead metrics are used.

Whilst a wide range of dataset variants have been explored and some are not presented due to space limitations, those included in the next section are representative of the findings of this research.

## 5. RESULTS

### 5.1 Feature Selection

In our experiments we have utilized the feature selection algorithms within Weka to identify which metrics in the set are considered significant. The CfsSubset Evaluator and Information Gain feature selection algorithms within Weka were applied to each dataset. The results are presented in Table 2.

**Table 2. Feature Selection Results**

| ID | Features Selected | |
|---|---|---|
| | Infogain | CfsSubset |
| 1 | 3, 30, 37, 14, 38, 16, 41, 7 | 3, 14, 16, 37 |
| 1a | 3, 7 | 3 |
| 1b | 14, 16 | 14, 16 |
| 1c | 30, 37, 38, 41 | 30, 37 |
| 1d | 30, 37, 14, 38, 16, 41 | 14, 16, 30, 37 |
| 2 | 9, 2, 23, 11, 33, 32, 14, 40, 28, 27, 1, 16, 8, 29 Omitted: 42 | 2, 8, 9, 11, 14, 23, 27, 28, 33 |
| 2a | 9, 2, 11, 1, 8 | 2, 8, 9, 11 |
| 2b | 23, 14, 28, 27, 16, 29 | 14, 23, 27 , 28, 29 |
| 2c | 33, 32, 40, 42 | 32, 33, 42 |
| 2d | 9, 23, 11, 33, 32, 14, 40, 28 , 27, 1, 16, 8, 29. Omitted: 42 | 8, 9, 11, 23, 27, 28 , 32, 33 |
| 3 | 8, 1, 11, 35, 24, 20, 19, 14, 10, 33 | 1, 8, 10, 11, 33, 35 |
| 3a | 8, 1, 11, 10 | 1, 8, 10, 11 |
| 3b | 24, 19, 20, 14 | 14, 19, 20, 24 |
| 3c | 1, 35, 33 | 1, 33, 35 |
| 3d | 8, 1, 11, 35, 24, 19, 20, 14, 10, 33 | 1, 8, 10, 11, 33, 35 |

The histogram shown in Figure 1 indicates the frequency of selection for each of the metrics under consideration. In some cases, metrics were omitted from the experiments if their significance was low. There is a clear indication that certain metrics are insignificant as they are not selected at all, irrespective of the feature selection algorithm used. Similarly, metrics such as the average number of attributes per class, comment to code ratio, number of import statements, number of methods and the number of unique operators were selected very frequently, suggesting that they are stronger "code quality" indicators for the prediction of either build failure or success.

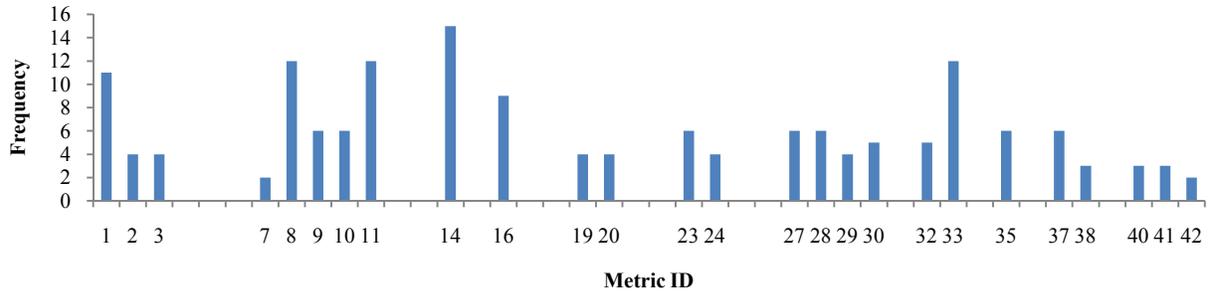

**Figure 1. Metric Selection Frequency**

## 5.2 Classification Results (InfoGain)

For each of the datasets we have used the J48 classification algorithm to attempt to discover common patterns amongst the metrics. Given the relatively small size of the data set we utilized 10-fold cross validation in order to make the best use of the training data. We acknowledge the relative optimism of cross validation and will address this in future work when more data becomes available from the Jazz project.

Table 3 shows the accuracy of the classification for each dataset with the features selected using the InfoGain feature selection algorithm in Weka. The overall accuracy is given in each case along with the number of correctly (and incorrectly) classified builds. The bracketed values refer to the number falsely predicted to be either failures (in the case of the "Failed Builds" column) or successes (in the case of the "Successful Builds") column.

**Table 3. Classification Results using InfoGain**

| ID | Accuracy | # Failed Builds Correct(Incorrect) | # Successful Builds Correct(Incorrect) |
|---|---|---|---|
| 1 | 68.2171 % | 22 (29 ) | 66(12 ) |
| 1a | 71.3178% | 23(28) | 69(9) |
| 1b | 69.7674% | 28(23) | 62(16) |
| 1c | 67.4419% | 15(36) | 72(6) |
| 1d | 75.1938% | 28(23) | 69(9) |
| 2 | 80.6202 % | 37(14) | 67(11) |
| 2a | 74.4186% | 26(25) | 70(8) |
| 2b | 73.6434% | 31(20) | 64(14) |
| 2c | 72.09 % | 23(28) | 70(8) |
| 2d | 76.7442% | 36(15) | 63(15) |
| 3 | 76.7442% | 34(17) | 65(13) |
| 3a | 58.1395% | 28(23) | 47(31) |
| 3b | 71.3178% | 32(19) | 60(18) |
| 3c | 65.1163% | 19(32) | 65(13) |
| 3d | 76.7442% | 34(17) | 13(65) |

From the results it is clear that the prediction of failed builds is generally more challenging than the classification of successful builds. The best classification result is for dataset 2, namely the use of the maximum values of each metric found in the source code associated with the build.

It is unsurprising that the maximum value dataset yields the best result. Generally, extremes of values for each metric are likely to be either desirable or undesirable, depending on the metric, and have the effect of being able to find "good" or "bad" code that will contribute towards either build failure or success. For example if the maximum value of the comment to code ratio taken across all work items exceeds a certain threshold then one would expect the build to be vulnerable to failure. This is demonstrated in Figure 2.

Inspection of the classification tree indicates that generally the first few nodes are intuitive. A Comment/Code Ratio greater than ~115 (which indicates 1 comment per 115 lines of code) is associated with build failure. A lack of commenting in code can lead to poor understandability. However a greater degree of commenting does not guarantee build success as when lack of cohesion is higher than 0.98 (which implies the system has no cohesion) that there is still a chance that a build may fail. That chance of failure depends on the number of attributes per class, essentially an implicit measure of both size and complexity.

Below these initial nodes there are ranges of values and repeated metrics indicating some confusion in the classification. Resolution of this uncertainty requires further research, however it may be related to the use of the maximum metric values in the data set.

## 5.3 Classification Results (CfsSubset)

Table 4 shows the results obtained for each dataset when the metrics chosen have been identified using the CfsSubset feature selection algorithm. A key difference between CfsSubset and InfoGain is that the CfsSubset algorithm takes into account combinations of features and not features in isolation.

**Table 4. Classification Results using CfsSubset**

| ID | Accuracy | # Failed Builds Correct(Incorrect) | # Successful Builds Correct(Incorrect) |
|---|---|---|---|
| 1 | 68.9922% | 23(28) | 66 (12 ) |
| 1a | 67.4419% | 28(23) | 59(19) |
| 1b | 69.7674% | 28(23) | 62(16) |
| 1c | 67.4419% | 15(36) | 72(6) |
| 1d | 75.969% | 29(22) | 69(9) |
| 2 | 75.1938 % | 32(19) | 65(13) |
| 2a | 79.0698% | 27(24) | 75(3) |
| 2b | 72.8682% | 30(21) | 64(14) |
| 2c | 71.3178% | 23(28) | 69(9) |
| 2d | 75.969 % | 31(20) | 67(11) |
| 3 | 64.3411% | 22(29) | 61(17) |
| 3a | 58.9147% | 29(22) | 47(31) |
| 3b | 71.3178% | 32(19) | 60(18) |
| 3c | 65.1163% | 19(32) | 65(13) |
| 3d | 64.3411% | 22(29) | 61(17) |

As with the results found using the InfoGain feature selection algorithm, it appears more challenging to identify causes of failure. A key difference in the results is that the overall accuracy can be traded off against an improved ability to predict failed builds. For example, using dataset 2a gives a very large number of correctly classified successful builds that leads to the highest overall accuracy. Meanwhile, using dataset 3b gives a more accurate classification of failed builds at the cost of a reduced ability to classify successful build. Given the challenge in identifying failed builds the best classification is again with dataset 2, which provided the best tradeoff between overall accuracy and correct classification of failed builds. The model for the best result is given in Figure 3.

Inspection of the classification tree shows a large overlap, in the higher nodes at least, with the previous results. However there is a greater deal of clarity in the lower nodes in the tree. The only repeated metrics arise when there is a lack of cohesion > 0.98, with an average number of attributes per class of <= 52.5 and an average block depth of >8. Beyond that there is a degree of confusion on this branch of the classification tree; however all other branches have a degree of intuitiveness.

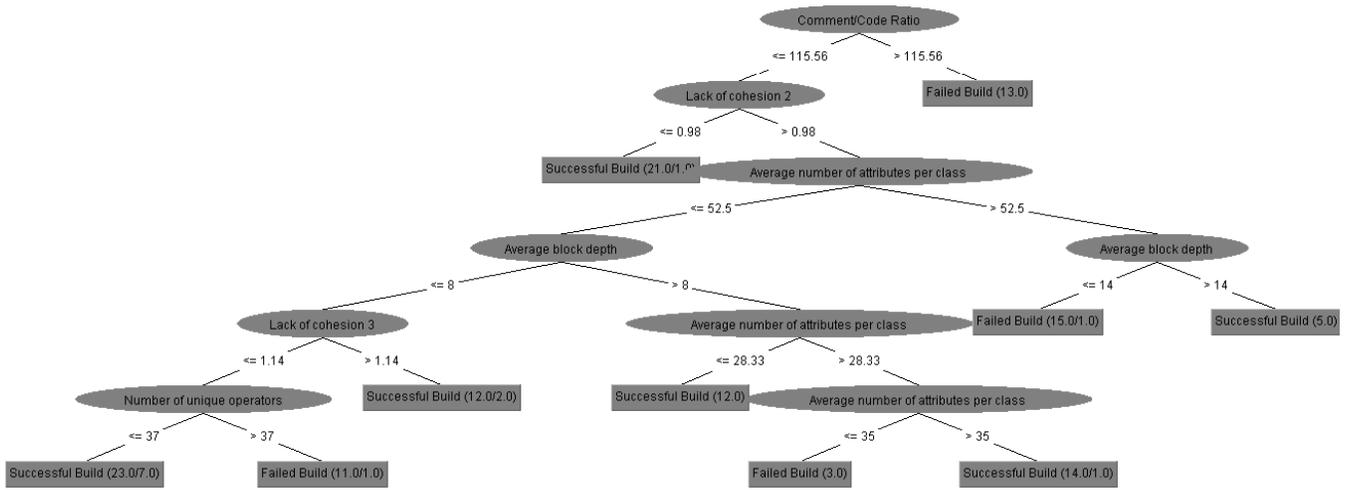

**Figure 2. Classification Tree (CfsSubSet Selection)**

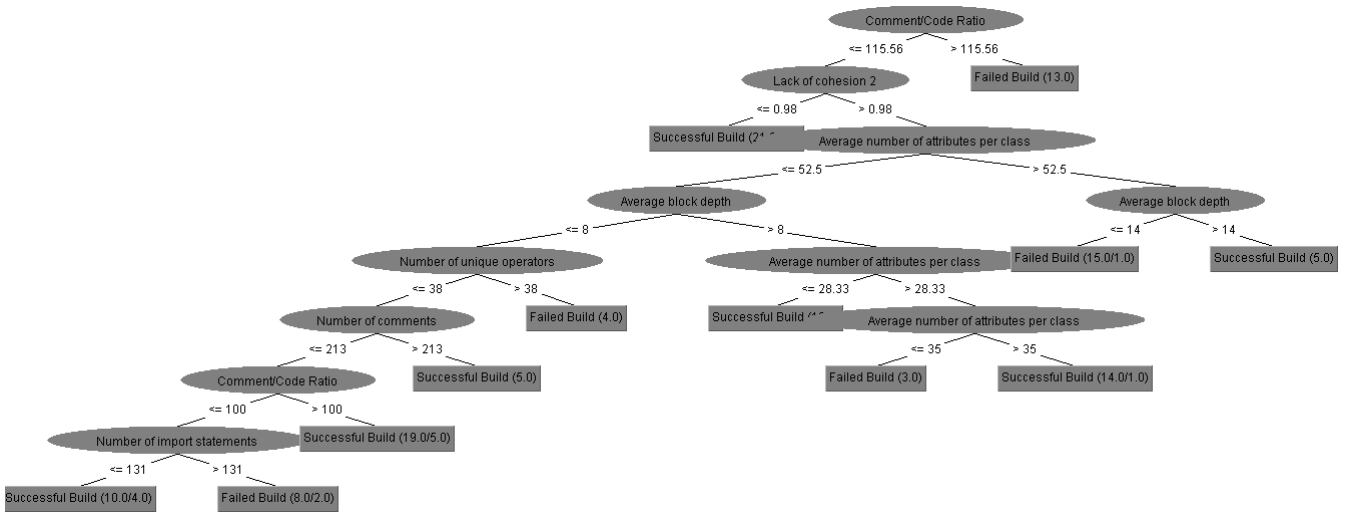

**Figure 3. Classification Tree (InfoGain Selection)**

## 5.4 Classification Results (Frequency)

There is a relatively high degree of variability in the features selected from each of the datasets presented in section 5.1 and the resulting classification experiments in sections 5.2 and 5.3. Therefore a third set of experiments were conducted where the metrics selected for inclusion are filtered using a different approach. From Figure 1 a number of metric sets have been chosen based on the frequency of selection across different datasets. These most frequently selected metrics have been applied to the three base datasets shown in Table 1.

The metric sets have been chosen by applying a threshold on the frequency of selection with any of the two feature selection algorithms applied to any given dataset. The threshold was varied from 4 to 10 in increments of 2. The selected features are show in Table 5.

**Table 5. Selected Metrics by Frequency**

| ID | Frequency Threshold | Selected Metrics |
|---|---|---|
| A | 4+ | 1, 8, 9, 10, 11, 14, 16, 23, 27, 28, 30, 32, 33, 35, 37 |
| B | 6+ | 1, 8, 9, 10, 11, 14, 16, 23, 27, 28, 33, 35, 37 |
| C | 8+ | 1, 8, 11, 14, 16, 33 |
| D | 10+ | 1, 8, 11, 14, 33 |

Table 6 shows the classification results with these selected metrics on the three main datasets. The ID shows the base dataset ID with the ID of metrics selected on the basis of their frequency of occurrence.

**Table 6. Classification Results using Frequency**

| ID | Accuracy | # Failed Builds Correct(Incorrect) | # Successful Builds Correct(Incorrect) |
|---|---|---|---|
| 1A | 67.4419% | 30(21) | 57(21) |
| 1B | 68.9922% | 29(22) | 60(18) |
| 1C | 71.3178% | 26(25) | 66(12) |
| 1D | 67.4419% | 17(34) | 70(8) |
| 2A | 75.1938% | 33(18) | 64(14) |
| 2B | 75.1938 % | 33(18) | 64(14) |
| 2C | 74.4186% | 26(25) | 70(8) |
| 2D | 74.4186% | 28(23) | 68(10) |
| 3A | 72.8682% | 30(21) | 64(14) |
| 3B | 72.093 % | 31(20) | 62(16) |
| 3C | 67.4419% | 25(26) | 62(16) |
| 3D | 74.4186% | 28(23) | 68(10) |

As with the other experiments it appears that there is more challenge in correctly classifying failed builds. The best results occur with the maximum value dataset, with identical results for metrics selected at least 4 and 6 times by the previous feature selection experiments.

Interestingly, as the number of selected metrics is reduced by applying a higher frequency threshold, the overall accuracy does not change significantly, yet there is a trend towards better classification of successful builds. This possibly indicates that some metrics are very strong indicators of success whereas others are weak indicators of failure.

The classification tree for experiment 2A is shown in Figure 4.

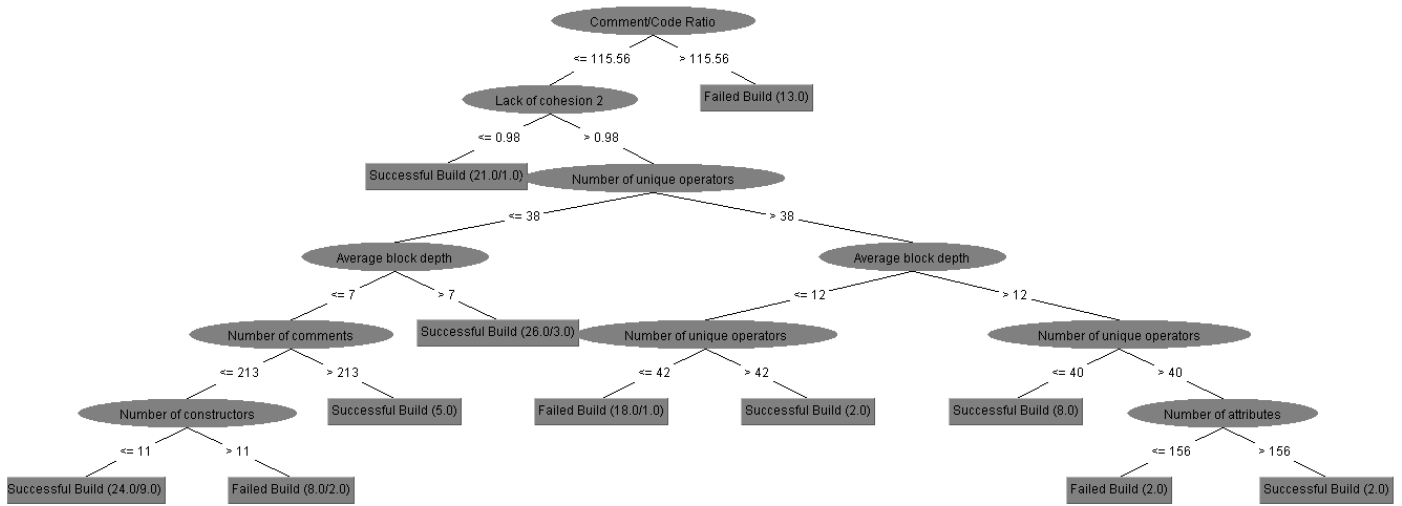

**Figure 4. Classification Tree (Frequency Selection)**

This classification tree shares the same top level nodes as those shown in Figure 2 and Figure 3, namely the first indicators are Comment/Code Ratio and Lack of Cohesion. When the Lack of Cohesion is > 0.98 the next indicator is the Number of Unique Operators. When this is <=38, there is no confusion in the lower branch of the classification tree. Failed builds occur when the Average Block Depth is small; there are relatively few comments and a high number of constructors.

For the other branch, when the Number of Unique Operators is > 38 there is still a degree of confusion with in the branches of the tree. It appears that a small number of instances are influencing the classification and potentially analyzing these is a source of removing the confusion from the tree. For example, when the Number of Unique Operators is > 38 and the Average Block Depth > 12, there are 8 successful builds where the Number of Unique Operators is either 39 or 40. Similarly, there are only 2 successful build where the Number of Unique Operators is > 42. The degree of confusion is in fact a very narrow slice which is being created by a very small number of instances of builds that have unique characteristics that are influencing the classification.

## 6. LIMITATIONS & FURTHER WORK

Most of the limitations in the current study are products of the relatively small sample size of build data from the Jazz project combined with the sparseness of the data itself. For example, the ratio of metrics (48) to builds (120) is such that it is difficult to truly identify significant metrics. As a result, the need to select maximum values of metric values biases the classification and probably causes some of the confusion in the classification trees.

Whilst a new release of the Jazz repository is pending, in the meantime the main thrust of our future work is to expand the build data to improve the degree of granularity and potentially remove some of the confusion in the classification trees.

Another key aspect for further study is to investigate why predicting failures is harder than predicting successes. In particular, we have observed that predicting failure is a different task than predicting non-success. This is due to the fact that the build successes and failures overlap in feature space and "failure" signatures have a greater degree of fragmentation than their "success" counterparts. As a result, the final aspect of future work is to develop a deeper understanding of what source code characteristics are most related to build failure and develop a set of 'best practice' guidelines for software development.

## 7. CONCLUSIONS

This paper presents the outcomes of an initial systematic attempt to predict build success and/or failure for a software product by utilizing source code metrics. Prediction accuracies of 70-80% have been achieved through the use of the J48 classification algorithm combined with 10-fold cross validation. Despite this high overall accuracy, there is greater difficulty in predicting failure than success and at present the classification trees content some uncertainty and confusion, but show promise in terms of informing software development activities in order to minimize the chance of failure.

## 8. ACKNOWLEDGMENTS

Our thanks go to IBM for providing access to the Jazz repository and the BuildIT consortium that has funded this research. We also like to thank Professor Stephen MacDonell from Auckland University of Technology for providing valuable expertise regarding software metrics.